\documentclass[prb,showpacs,amsmath,amssymb]{revtex4}
\usepackage{graphicx}
\usepackage{bm}

\def\br{\bm{r}}
\def\bk{\bm{k}}
\def\bq{\bm{q}}
\def\bQ{\bm{Q}}
\def\im{\,\mathrm{Im}}

\def\Tr{\mathrm{Tr}\,}
\def\det{\mathrm{det}}
\def\tr{\mathrm{tr}\,}
\def\beps{\bm{\varepsilon}}
\def\bvs{\bm{v}_s}

\begin{document}

\title{Goldstone modes in Larkin-Ovchinnikov-Fulde-Ferrell superconductors}

\author{K. V. Samokhin}

\affiliation{Department of Physics, Brock University,
St.Catharines, Ontario, Canada L2S 3A1}
\date{\today}

\begin{abstract}
In nonuniform Larkin-Ovchinnikov-Fulde-Ferrell (LOFF) superconductors, both the gauge symmetry and the continuous translational symmetry of the normal state are spontaneously broken. This leads to additional bosonic excitations, or Goldstone modes, corresponding to the deformations of the order parameter amplitude modulation in real space. We derive general expressions for the energy of the phase and elastic Goldstone modes. As an example, the superfluid density and the elastic modulus of a one-dimensional LOFF superconductor are calculated at low temperatures. 
\end{abstract}

\pacs{74.20.-z, 74.25.Ha, 74.40.-n}

\maketitle

\section{Introduction}
\label{sec: intro}

In most superconductors, the Cooper pairing is suppressed by the coupling of the orbital motion of electrons with a magnetic field.\cite{Tink96}
In some cases, however, the orbital suppression is ineffective and the dominant mechanism of the magnetic pair breaking is spin-related, caused by either the Zeeman
coupling with the external field, or the exchange interaction with the localized spins in a magnetic crystal. The superconductivity is then said to be paramagnetically limited. It was shown by Larkin and Ovchinnikov\cite{LO64} (LO) and Fulde and Ferrell\cite{FF64} (FF) that
the competition between the paramagnetic pair breaking and the condensation energy results in the formation
of a peculiar nonuniform superconducting state -- the LOFF state -- in which the Cooper pairs have nonzero center-of-mass momenta. In the simplest cases, the order parameter is described by either a single plane wave: $\Delta(\br)=\Delta_0e^{i\bQ\br}$ (FF state), or a superposition of two plane waves: $\Delta(\br)=\Delta_0\cos\bQ\br$ (LO state). In general, minimization of the free energy yields more complicated structures.

The experimental detection of the LOFF state requires both the orbital effects and disorder to be sufficiently weak. This can be achieved in the superconductors with a short coherence length and heavy effective mass of quasiparticles, which increases the orbital upper critical field $H_{c2}$, and can be particularly favored in the systems with low-dimensional geometry, either in real space (superconductivity in films or at interfaces, in a parallel field) or in momentum space (quasi-one-dimensional and quasi-two-dimensional band structures). For a review of the LOFF state in superconductors, see Ref. \onlinecite{FFLO-review-07}. A different route to the experimental realization of the LOFF state is offered by cold atomic Fermi gases. In these systems the role of the Zeeman field is played by the mismatch between the Fermi surfaces, which is controlled by the difference between the numbers of atoms in two different hyperfine states, see Ref. \onlinecite{RS09} for a review. Further afield, states similar to the LOFF state have been discussed recently in high-energy physics, in the context of ``color superconducting'' quark matter.\cite{CasNar04}
While most theoretical studies have focused on the equilibrium properties of the LOFF systems, in particular, the phase diagram and the most stable order parameter, a number of recent works have looked at deviations from a perfectly periodic order, including thermal and quantum fluctuations\cite{Shima98,Ohashi02,SM06,KCB07} and topological defects, such as vortices and dislocations.\cite{Shima98,AMZ08,AT08} 

In addition to the $U(1)$ phase rotation symmetry, the LOFF state can break the continuous translational symmetry, which leads to a richer spectrum of low-energy bosonic excitations, or Goldstone modes. Since neither global phase rotations nor uniform translations of the order parameter cost any energy, the Goldstone modes in the LOFF state are associated with slow gradients of the order parameter phase, as well as with weakly nonuniform displacements of the order parameter amplitude modulation. In the lowest order, the energy is quadratic in both the phase and the displacement gradients. For instance, one can expect that the energy of long-wavelength static phase fluctuations has the usual form $\rho_sv_s^2/2$, where the superfluid velocity $\bvs$ is proportional to the gradient of the fluctuating phase, and $\rho_s$ is the superfluid density, which characterizes the phase stiffness of the superconducting state.

In this paper, we present a microscopic derivation of the free energy of the Goldstone modes in a LOFF superconductor. Recently, a similar problem has been considered using the Ginzburg-Landau expansion near the critical temperature in the isotropic case, which is applicable to either a Fermi gas with a population imbalance,\cite{RV09} or the quark matter.\cite{MRS07} In Ref. \onlinecite{Ikeda07}, the elastic moduli of a LOFF vortex lattice were calculated. In contrast to these works, which used the free energy expansion in powers of the order parameter, we employ a transformation of the slow deformations of the LOFF order parameter into small perturbations in the effective bosonic action. General expressions for the effective action of the static Goldstone modes, both in the FF and LO states, are derived in Sec. \ref{sec: general}. In Sec. \ref{sec: 1D}, we illustrate how the formalism works using as an example the LO phase in a one-dimensional superconductor at low temperatures, for which the order parameter and the spectrum of excitations in the mean-field state are known exactly. Throughout the paper we use the units in which $\hbar=k_B=1$.

\section{Effective bosonic action in the LOFF state}
\label{sec: general}

We consider a spin-singlet superconductor without disorder, in an external magnetic field $\bm{B}$. The orbital pair breaking is neglected, so that the superconductivity is affected by the field only through the Zeeman splitting of the electron bands. The Hamiltonian is given by
\begin{eqnarray}
\label{H}
    H=\int d\br\,\psi_\alpha^\dagger(\br)(\hat\xi\delta_{\alpha\beta}+h\sigma_{3,\alpha\beta})\psi_\beta(\br)
    -V\int d\br\,\psi^\dagger_\uparrow(\br)\psi^\dagger_\downarrow(\br)\psi_\downarrow(\br)\psi_\uparrow(\br).
\end{eqnarray}
In the first term, $\hat\xi=\xi(\hat{\bk})$ is the effective band Hamiltonian (which includes the chemical potential $\mu$), $\hat{\bk}=-i\bm{\nabla}$, $\alpha,\beta=\uparrow,\downarrow$ is the spin projection on the
quantization axis along $\bm{B}$, $h=(g/2)\mu_B B$ is the Zeeman field (the electron charge is equal to $-e$), $\mu_B$ is the Bohr magneton, $g$ is the Land\'e factor,
and $\hat\sigma_3$ is the Pauli matrix. The electron wavefunctions are assumed to satisfy the periodic boundary conditions in a cubic box of side $L$ and volume 
${\cal V}=L^3$. The second term in Eq. (\ref{H}) is the Bardeen-Cooper-Schrieffer (BCS) pairing interaction with the coupling constant $V>0$. The Hamiltonian (\ref{H}) can also be applied to a ferromagnetic superconductor in zero applied field, in which
the electron bands are split due to the exchange interaction with the magnetically ordered localized spins, or to an imbalanced atomic Fermi gas.

We derive the free energy of our superconductor starting with the standard expression for the partition function in terms of the Grassmann functional integral:
$Z=\int{\cal D}\psi_\alpha{\cal D}\bar\psi_\alpha\,e^{-S}$, see, e.g. Ref. \onlinecite{Popov91}. Here $\psi_\alpha(\br,\tau)$ and $\bar\psi_\alpha(\br,\tau)$ are fermionic fields, $S=\int_0^\beta d\tau\left[\int d\br\,\bar\psi_\alpha\partial_\tau\psi_\alpha+H(\tau)\right]$ is the action associated with the Hamiltonian (\ref{H}), and $\beta=1/T$. 
Decoupling the interaction term by the Hubbard-Stratonovich transformation, the partition function can be written as a
functional integral over a complex bosonic field $\Delta(\br,\tau)$ which has the meaning of a fluctuating superconducting order parameter:
$Z=\int{\cal D}\Delta{\cal D}\Delta^*\,e^{-S_{eff}[\Delta]}$, where $S_{eff}$ is the effective bosonic action. Focusing on the static fluctuations, the effective action takes the following form: $S_{eff}=\beta{\cal F}[\Delta]$, where 
\begin{equation}
\label{F-eff}
	{\cal F}=-T\sum_n\Tr\ln\hat{\cal G}^{-1}+\frac{1}{V}\int d^3\br\,|\Delta(\br)|^2
\end{equation}
is the free energy, $\hat{\cal G}^{-1}=i\omega_n-\hat{\cal H}$, $\omega_n=(2n+1)\pi T$ is the fermionic Matsubara frequency, and
\begin{equation}
\label{cal H}
	\hat{\cal H}=\left(\begin{array}{cc}
        \hat\xi+h & \Delta(\br) \\
        \Delta^*(\br) & -\hat\xi+h \\
    \end{array}\right)
\end{equation}
is the Bogoliubov-de Gennes (BdG) Hamiltonian. The trace in the first term in Eq. (\ref{F-eff}) is taken in both the coordinate and the electron-hole spaces and can be formally written as follows: $\Tr\ln\hat{\cal G}^{-1}=\sum_a\ln(i\omega_n-E_a)$, where $E_a$ are the eigenvalues of $\hat{\cal H}$, labelled by quantum numbers $a$. The sum over $a$ formally diverges and should be regularized by subtracting the corresponding expression in the normal state (in the spirit of the BCS model, we assume that  only a finite number of the electron states in a narrow energy interval near the Fermi energy are affected by superconductivity). Since we are interested in the energy of fluctuations, which vanishes in the normal state, we will not write the regularizing terms explicitly. 

The mean-field order parameter, denoted by $\Delta_0(\br)$, corresponds to a saddle point of the effective action, which is found from the equation
$\delta{\cal F}/\delta\Delta^*=0$. The solution is sensitive to the electronic band structure and the pairing symmetry. The problem of finding the most stable LOFF state at all $T$ and $h$ has not been solved even in the simplest cases, apart from an exact solution in one dimension, see Refs. \onlinecite{MN84} and \onlinecite{BP87}. In a fully isotropic three-dimensional system, it was argued in Ref. \onlinecite{MC04} that the phase transition between the normal and the LOFF states is always first order, and that the order parameter is represented by a sum of one, two, or three cosines, as the temperature is lowered.
In this work, we consider only the phases with a one-dimensional (1D) periodicity, see Appendix \ref{sec: Goldstones}: the FF phase with
\begin{equation}
\label{FF-mean field}
	\Delta_{0,FF}(\br)=\Delta_0 e^{iQz},
\end{equation}
and the LO phase with
\begin{equation}
\label{LO-mean field}
	\Delta_{0,LO}(\br)=\Delta_0 f(z),
\end{equation}
where $f(z)$ is a real periodic function with the period $d$. In the vicinity of a second-order normal metal-LOFF superconductor phase transition, one can put $f(z)=\cos Qz$, where $Q=2\pi/d$. Away from the critical temperature, nonlinear effects add higher Fourier harmonics to the order parameter:
\begin{equation}
\label{f-z-fourier}
	 f(z)=\sum_{p=-\infty}^\infty f_p e^{ipQz},\quad f_{-p}=f_p^*.
\end{equation}
The sum here does not include the $p=0$ harmonic, so that the spatial average of the order parameter is zero. Thus, the nonuniform superconducting state resembles a periodic array of domain walls separating the regions where the order parameter is almost uniform.\cite{BR94}

\subsection{FF phase}
\label{sec: FF}

The mean-field solutions described above are not unique. While one can uniformly rotate the overall phase of the order parameter without
any energy penalty, a nonuniform phase rotation corresponds to a Goldstone mode, whose energy is small in the long-wavelength limit. In contrast, even uniform deviations of the order parameter amplitude $\Delta_0$ from its mean-field value cost energy and can be neglected. Thus, the low-energy fluctuations of the order parameter in the FF phase are described by the following expression:
\begin{equation}
\label{deformed FF}
	\Delta_{FF}(\br)=\Delta_0 e^{i\theta(\br)} e^{iQz},
\end{equation}
where $\theta$ varies slowly on the scale of the LOFF period $d$. The fluctuating phase $\theta$ can be removed from the off-diagonal elements of the BdG Hamiltonian (\ref{cal H}) by the unitary transformation $\hat{\cal H}\to{\hat{\tilde{\cal H}}}=\exp(-i\theta\hat\sigma_3/2)\hat{\cal H}\exp(i\theta\hat\sigma_3/2)$, and we obtain:
\begin{equation}
\label{H_FF-gen}
	{\hat{\tilde{\cal H}}}=\left(\begin{array}{cc}
        \xi(\hat{\bm{\pi}}_+) +h & \Delta_0 e^{iQz} \\
        \Delta_0 e^{-iQz} & -\xi(\hat{\bm{\pi}}_-) +h \\
    \end{array}\right),
\end{equation}
where $\hat{\bm{\pi}}_\pm=\hat{\bk}\pm m\bvs$, $\bvs=\bm{\nabla}\theta/2m$ is the superfluid velocity, and $m$ is the electron mass.

For slow fluctuations, one can use a constant superfluid velocity and expand the BdG Hamiltonian (\ref{H_FF-gen}) in powers of $\bvs$. For the lowest two orders in the perturbation series we obtain: ${\hat{\tilde{\cal H}}}=\hat{\cal H}_{0,FF}+\delta\hat{\cal H}$, where
\begin{equation}
\label{cal H_0}
	\hat{\cal H}_{0,FF}=\left(\begin{array}{cc}
        \hat\xi+h & \Delta_0 e^{iQz} \\
        \Delta_0 e^{-iQz} & -\hat\xi+h \\
    \end{array}\right)
\end{equation}
is the mean-field Hamiltonian,
\begin{equation}
\label{delta H-FF}
	\delta\hat{\cal H}=mv_{s,i}v_i(\hat{\bk})\hat\sigma_0+\frac{1}{2}m^2v_{s,i}v_{s,j}m_{ij}^{-1}(\hat{\bk})\hat\sigma_3,
\end{equation}
$\bm{v}=\partial\xi/\partial\bk$ is the quasiparticle velocity, and $m_{ij}^{-1}=\partial^2\xi/\partial k_i\partial k_j$ is the inverse tensor of effective masses.

Using Eq. (\ref{F-eff}), the phase fluctuation contribution to the free energy can be written as follows: 
$$
	\delta{\cal F}_{FF}=-T\sum_n\bigl[\Tr\ln(\hat{\cal G}_0^{-1}-\delta\hat{\cal H})-\Tr\ln\hat{\cal G}_0^{-1}\bigr]
	=T\sum_n\left[\Tr(\hat{\cal G}_0\delta\hat{\cal H})+\frac{1}{2}\Tr(\hat{\cal G}_0\delta\hat{\cal H}\,\hat{\cal G}_0\delta\hat{\cal H})+...\right],
$$
where $\hat{\cal G}_0=(i\omega_n-\hat{\cal H}_{0,FF})^{-1}$ is the $2\times 2$ matrix Green's function in the undeformed FF state. Keeping only the quadratic terms in $\bvs$ (the first order terms vanish at the saddle point), we obtain the energy of the phase Goldstone modes per unit volume:
\begin{equation}
\label{F_FF}
	\frac{\delta{\cal F}_{FF}}{\cal V}=\frac{1}{2}\rho_{s,ij}v_{s,i}v_{s,j},
\end{equation}
where
\begin{equation}
\label{rho_s_ij}
	\rho_{s,ij}=m^2\frac{1}{\cal V}T\sum_n\bigl[\Tr(\hat v_i\hat{\cal G}_0\hat v_j\hat{\cal G}_0)
	+\Tr(\hat m_{ij}^{-1}\hat\sigma_3\hat{\cal G}_0)\bigr]
\end{equation}
is the superfluid mass density tensor. In the normal state, the Green's function is electron-hole diagonal, translationally-invariant, and satisfies the identity 
\begin{equation}
\label{Ward-ident}
	\frac{\partial\hat{\cal G}_0(\bk,\omega_n)}{\partial\bk}=\bm{v}(\bk)\hat\sigma_3\hat{\cal G}_0^2(\bk,\omega_n).	
\end{equation}
Using $m^{-1}_{ij}=\partial v_i/\partial k_j$ and integrating the second term in Eq. (\ref{rho_s_ij}) by parts, one can verify that in the normal state the superfluid density vanishes.

\subsection{LO phase}
\label{sec: LO}

In addition to the phase rotation symmetry, the LOFF order parameter can also break continuous translational symmetry, leading to the existence of additional Goldstone modes, see Appendix \ref{sec: Goldstones}. In the LO phase, these additional modes correspond to weakly nonuniform deformations of the amplitude modulation. 

Suppose a Cooper pair that was at a point $\br'$ in the undeformed state is found at a point $\br$ after the deformation characterized by the displacement $\bm{u}(\br')=\br-\br'$ (we assume that the relation between $\br$ and $\br'$ is unique and invertible). As a result, the pair wavefunction $\Delta_0(\br')$ is transported to the new location, leading to the order parameter at the point $\br$ being transformed from $\Delta_0(\br)$ into $\Delta(\br)=\Delta_0(\br'(\br))$.  For the LO phase with a 1D periodicity, described by Eq. (\ref{LO-mean field}), the deformed order parameter has the following form:
\begin{equation}
\label{deformed LO}
	\Delta_{LO}(\br)=\Delta_0 e^{i\theta(\br'(\br))}f(z'(\br)),
\end{equation}
where the relation between $\br$ and $\br'$ is defined by the expressions $x=x'$, $y=y'$, $z=z'+u(x',y',z')$, and both $\theta$ and $u$ vary slowly compared with $d$. Due to the periodicity conditions for the wavefunctions, the region ${\cal R}$ in three-dimensional Euclidean space occupied by the superconductor is homeomorphic to a three-dimensional torus. The ``Eulerian'' coordinates $\br$ and the ``Lagrangian'' coordinates $\br'$ (Ref. \onlinecite{ChLub-book}) correspond to different parametrizations of the points in ${\cal R}$. Since the order parameter also has to be periodic, we assume that $u(x'+L,y',z')=u(x',y',z')$, \textit{etc}, therefore, $x'(x+L,y,z)-x'(x,y,z)=L$, \textit{etc}.
Since neither a global phase rotation, described by a constant $\theta$, nor a uniform translation, described by a constant $u$, cost any energy, the free energy ${\cal F}$ can only depend on the gradients of $\theta$ and $u$.

It is convenient to transform the BdG Hamiltonian of the LO phase into the coordinate system in which the order parameter is not deformed. This can be achieved by changing coordinates from $\br$ to $\br'$, which affects the metric of the Euclidean space.\cite{DFN84} In the Lagrangian coordinates, Eq. (\ref{cal H}) takes the form
\begin{equation}
\label{H-prime}
	\hat{\cal H}'=\left(\begin{array}{cc}
        \hat\xi'+h & \Delta_0 e^{i\theta(\br')} f(z') \\
        \Delta_0 e^{-i\theta(\br')} f(z') & -\hat\xi' +h \\
    \end{array}\right),	
\end{equation}
while the components of the metric tensor become
\begin{equation}
\label{g-ij}
	g_{ij}(\br')=\left(\delta_{ki}+\delta_{k3}\frac{\partial u}{\partial x^{\prime,i}}\right)
	\left(\delta_{kj}+\delta_{k3}\frac{\partial u}{\partial x^{\prime,j}}\right),
\end{equation}
where we introduced the notations $(x^1,x^2,x^3)\equiv(x,y,z)$ and $(x^{\prime,1},x^{\prime,2},x^{\prime,3})\equiv(x',y',z')$, and the repeated indices are summed over.
In order to find how the band Hamiltonian is transformed, we assume that it can be represented as a series expansion:
$$
	\hat\xi=\sum_{\mathrm{even\ }N}A^{i_1...i_N}\hat k_{i_1}...\hat k_{i_N}-\mu,
$$
where $A^{i_1...i_N}$ are determined by the crystal symmetry and completely symmetric with respect to the permutation of $i_1,...,i_N=1,2,3$. Under the change of coordinates $\br\to\br'$, this is transformed into
\begin{equation}
\label{xi prime}
	\hat\xi'=\sum_N A^{i_1...i_N}e^{j_1}_{i_1}\hat k'_{j_1}...e^{j_N}_{i_N}\hat k'_{j_N}-\mu,
\end{equation}
where $\hat k'_i=-i\nabla'_i$ and $e^j_i(\br')=\partial x^{\prime,j}/\partial x^i$. For instance, if the band dispersion can be treated in the effective mass approximation, i.e. $\xi(\bk)=\bk^2/2m^*-\mu$, then $\hat\xi'=-\Delta_{LB}/2m^*-\mu$, where
$\Delta_{LB}=g^{-1/2}\nabla'_ig^{1/2}g^{ij}\nabla'_j$
is the Laplace-Beltrami operator (Ref. \onlinecite{DFN84}), $g=(1+\partial u/\partial z')^2$ is the determinant of the metric tensor, and $g^{ij}$ are the components of the inverse of the metric tensor.

The partition function can be written as follows: $Z\propto\int{\cal D}\Delta{\cal D}\Delta^*\, e^{-\beta({\cal F}_1+{\cal F}_2)}$, where
\begin{equation}
\label{cal F_1}
	e^{-\beta{\cal F}_1}=\int {\cal D}\Psi{\cal D}\bar\Psi\,e^{\Tr(\bar\Psi\hat{\cal G}^{-1}\Psi)},
\end{equation}
$\Psi=(\psi_\uparrow,\bar\psi_\downarrow)^T$ and $\bar\Psi=(\bar\psi_\uparrow,\psi_\downarrow)$ are the Nambu Grassmann fields, ${\cal F}_2=V^{-1}
\int d^3\br\,|\Delta_{LO}(\br)|^2$, and $\Delta_{LO}$ is given by Eq. (\ref{deformed LO}).
One can show that ${\cal F}_2$ does not depend on the deformation. Indeed, using Eqs. (\ref{LO-mean field}) and (\ref{f-z-fourier}), we obtain:
$$
	{\cal F}_2=\frac{1}{V}\int d^3\br'\sqrt{g}\,|\Delta_{0,LO}(\br')|^2=\frac{\Delta_0^2}{V}\sum_{p_1p_2}I_{p_1p_2}f_{p_1}f_{p_2}.
$$
Here $I_{p_1p_2}=\int d^3\br'\sqrt{g}\,e^{i(p_1+p_2)Qz'}\simeq{\cal V}\delta_{p_1,-p_2}$, if the displacement $u$ varies on the scale much greater than the LOFF period. Therefore, ${\cal F}_2=({\cal V}\Delta_0^2/V)\sum_p|f_p|^2$, i.e. all the effects of the order parameter deformation are contained in ${\cal F}_1$.

Since the transformation from $\br$ to $\br'$ changes the scalar product in the functional space, one should be careful calculating the integral in Eq. (\ref{cal F_1}). Let us introduce a complete and orthonormal set of 
wavefunctions in ${\cal R}$, using the Eulerian coordinates $\br$. One can use, for instance, the eigenfunctions of $\hat\xi$, i.e. the normalized plane waves $\chi_{\bk}(\br)={\cal V}^{-1/2}e^{i\bk\br}$, where $\bk=(2\pi/L)(m_1,m_2,m_3)$, and $m_i$ take integer values due to the periodicity conditions. We construct the basis Nambu spinors labelled by $a=(\bk,s)$ ($s=1,2$ is the electron-hole index), as follows:
$\varphi_{\bk,1}=(\chi_{\bk},0)^T$ and $\varphi_{\bk,2}=(0,\chi_{\bk})^T$,
and represent the Nambu fields in the form
$$
    \Psi(\br,\tau)=T\sum_n\sum_a c_{n,a}\varphi_a(\br)e^{-i\omega_n\tau},\quad \bar\Psi(\br,\tau)=T\sum_n\sum_a \bar c_{n,a}\varphi^\dagger_a(\br)e^{i\omega_n\tau},
$$
where $c_{n,a}$ and $\bar c_{n,a}$ are elements of the Grassmann algebra. From Eq. (\ref{cal F_1}) we obtain:
\begin{equation}
\label{cal F_1-1}
	e^{-\beta{\cal F}_1}=\int\prod_n\prod_a dc_{n,a}d\bar c_{n,a}\, e^{\sum_{n,ab}{\cal G}_{ab}^{-1}(\omega_n)\bar c_{n,a}c_{n,b}}
	=\prod_n\det\,\hat{\cal G}^{-1}.
\end{equation}
The matrix elements of the inverse Green's operator are given by
$$
    	{\cal G}_{ab}^{-1}(\omega_n)=\int d^3\br\,\tr\varphi^\dagger_a(\br)(i\omega_n-\hat{\cal H})\varphi_b(\br)
	=\int d^3\br'\sqrt{g}\,\tr\varphi^\dagger_a(\br(\br'))(i\omega_n-\hat{\cal H}')\varphi_b(\br(\br')),
$$
where ``$\tr$'' denotes a $2\times 2$ matrix trace and $\hat{\cal H}'$ is given by Eq. (\ref{H-prime}). Introducing the new Nambu wavefunctions
\begin{equation}
\label{tilde-phis}
    	\tilde\varphi_a(\br')=g^{1/4}e^{-i\theta(\br')\hat\sigma_3/2}\varphi_a(\br(\br')),
\end{equation}
we obtain:
\begin{equation}
\label{Gab}
    	{\cal G}_{ab}^{-1}(\omega_n)=\int d^3\br'\,\tr\tilde\varphi_a^\dagger(\br')(i\omega_n-{\hat{\tilde{\cal H}}}) \tilde\varphi_b(\br'),
\end{equation}
where
\begin{equation}
\label{tilde-H}
	{\hat{\tilde{\cal H}}}=\left(\begin{array}{cc}
        \hat\xi'_+ +h & \Delta_0 f(z') \\
        \Delta_0 f(z') & -\hat\xi'_- +h \\
    \end{array}\right),
\end{equation}
and
\begin{equation}
\label{hat eps pm}
	\hat\xi'_\pm=e^{\mp i\theta/2}g^{1/4}\hat\xi'g^{-1/4}e^{\pm i\theta/2}
	=g^{1/4}\sum_N A^{i_1...i_N}e^{j_1}_{i_1}(\hat k'_{j_1}\pm mv_{s,j_1})...e^{j_N}_{i_N}(\hat k'_{j_N}\pm mv_{s,j_N})g^{-1/4}-\mu.
\end{equation}
The last expression follows from Eq. (\ref{xi prime}). 

The functions (\ref{tilde-phis}) form a complete and orthonormal set in ${\cal R}$. Since the factors $g^{1/4}$ have been absorbed into the definitions of $\tilde\varphi_a$, the integral in Eq. (\ref{Gab}) is formally the same as for fermions with the Hamiltonian ${\hat{\tilde{\cal H}}}$ moving in a flat space (with the metric tensor given by a unit matrix). Now one can drop the primes and write
$\tilde\varphi_a(\br)=\sum_b\varphi_b(\br)U_{ba}$, where $U_{ba}=\int d^3\br\,\tr\varphi^\dagger_b(\br)\tilde\varphi_a(\br)$ form a unitary matrix. Therefore,
using Eq. (\ref{cal F_1-1}), we obtain: $\det\,\hat{\cal G}^{-1}=\det\bigl[U^\dagger(i\omega_n-{\hat{\tilde{\cal H}}})U\bigr]=\det(i\omega_n-{\hat{\tilde{\cal H}}})$, and
\begin{equation}
\label{cal F_1-reg}
	{\cal F}_1=-T\sum_n\Tr\ln(i\omega_n-{\hat{\tilde{\cal H}}}).
\end{equation}
The advantage of using the transformed Hamiltonian ${\hat{\tilde{\cal H}}}$ is that, in contrast to $\hat{\cal H}$, it contains only the gradients of the fluctuating fields, which are small and can be treated perturbatively. 

We consider only the limit of a slow deformation, when the gradient of the displacement can be set to a constant: $\bm{\nabla}u=\beps$. Different components of $\beps$ play different roles: while $\varepsilon_{x,y}$ correspond to a uniform tilting, $\varepsilon_z$ describes a uniform compression or stretching of the order parameter modulation. The effective band Hamiltonian (\ref{xi prime}) is transformed into $\hat\xi=\xi(\hat\bk-\beps\hat k_z/(1+\varepsilon_z))$, while Eq. (\ref{hat eps pm}) takes the form
$\hat\xi_\pm=\xi(\hat{\bm{\pi}}_\pm-\beps\hat\pi_{\pm,z}/(1+\varepsilon_z))$, where $\hat{\bm{\pi}}_\pm=\hat{\bk}\pm m\bvs$ (recall that the primes denoting the Lagrangian coordinates have been dropped). Although the assumption of a constant strain is not consistent with the global periodicity conditions, it is legitimate if the scale of variation of $u$ is much greater than $d$, the period of the LOFF structure. This ``local limit'' fails when $d$ diverges, which happens, e.g., near the tricritical point on the $T-h$ phase diagram, where the Ginzburg-Landau gradient term changes sign. Similarly to the FF case, we also assume that the superfluid velocity $\bvs$ is uniform and small. 

For slow fluctuations, we expand the Hamiltonian (\ref{tilde-H}) in powers of $\bvs$ and $\beps$, and obtain: 
${\hat{\tilde{\cal H}}}=\hat{\cal H}_{0,LO}+\delta\hat{\cal H}$, where
\begin{equation}
\label{cal H_0-LO}
	\hat{\cal H}_{0,LO}=\left(\begin{array}{cc}
        \hat\xi+h & \Delta_0 f(z) \\
        \Delta_0 f(z) & -\hat\xi+h \\
    \end{array}\right),
\end{equation}
is the mean-field part, while
\begin{equation}
\label{delta H-LO}
	\delta\hat{\cal H}=(\hat\omega_i-\varepsilon_i\hat\omega_z)v_i(\hat{\bk})\hat\sigma_0
	+\frac{1}{2}\hat\omega_i\hat\omega_jm_{ij}^{-1}(\hat{\bk})\hat\sigma_3,
\end{equation}
with $\hat{\bm{\omega}}=m\bvs\hat\sigma_0-\beps\hat k_z\hat\sigma_3$, can be treated as a small perturbation. We keep only the lowest two orders in $\bvs$ and $\beps$.  The higher orders can be easily obtained from Eq. (\ref{hat eps pm}), if needed. 

The energy of the Goldstone modes follows from Eq. (\ref{cal F_1-reg}):
$$
	\delta{\cal F}_{LO}=T\sum_n\left[\Tr(\hat{\cal G}_0\delta\hat{\cal H})
		+\frac{1}{2}\Tr(\hat{\cal G}_0\delta\hat{\cal H}\,\hat{\cal G}_0\delta\hat{\cal H})+...\right],
$$
where $\hat{\cal G}_0=(i\omega_n-\hat{\cal H}_{0,LO})^{-1}$ the $2\times 2$ matrix Green's function in the undeformed LO state.
Retaining only the quadratic terms, we arrive at the following expression:
\begin{equation}
\label{E_LO}
	\frac{\delta{\cal F}_{LO}}{\cal V}=\frac{1}{2}\rho_{s,ij}v_{s,i}v_{s,j}+\frac{1}{2}K_{ij}\varepsilon_i\varepsilon_j
	+\tilde K_{ij}v_{s,i}\varepsilon_j.
\end{equation}
Here the superfluid density tensor is given by
\begin{equation}
\label{rho_s_ij-LO}
	\rho_{s,ij}=m^2\frac{1}{\cal V}T\sum_n\bigl[\Tr(\hat v_i\hat{\cal G}_0\hat v_j\hat{\cal G}_0)
	+\Tr(\hat m_{ij}^{-1}\hat\sigma_3\hat{\cal G}_0)\bigr],
\end{equation}
which has the same form as in the FF state, see Eq. (\ref{rho_s_ij}), but with a different Green's function,
\begin{equation}
\label{K_ij}
	K_{ij}=\frac{1}{\cal V}T\sum_n\bigl[\Tr(\hat v_i\hat k_z\hat\sigma_3\hat{\cal G}_0\hat v_j\hat k_z\hat\sigma_3\hat{\cal G}_0)
	+\Tr(\hat m_{ij}^{-1}\hat k_z^2\hat\sigma_3\hat{\cal G}_0)+2\delta_{iz}\Tr(\hat v_j\hat k_z\hat\sigma_3\hat{\cal G}_0)\bigr]
\end{equation}
is the tensor of elastic moduli, and
\begin{equation}
\label{tilde-K_ij}
	\tilde K_{ij}=-m\frac{1}{\cal V}T\sum_n\bigl[\Tr(\hat v_i\hat{\cal G}_0\hat v_j\hat k_z\hat\sigma_3\hat{\cal G}_0)
	+\Tr(\hat m_{ij}^{-1}\hat k_z\hat{\cal G}_0)+\delta_{iz}\Tr(\hat v_j\hat{\cal G}_0)\bigr].
\end{equation}
It is straightforward to show, using the identity (\ref{Ward-ident}), that all the stiffness coefficients, Eqs. (\ref{rho_s_ij-LO}), (\ref{K_ij}), and (\ref{tilde-K_ij}), vanish in the normal state. 

If the order parameter has a center of inversion, i.e. $\Delta(-\br)=\Delta(\br)$, then the Green's function satisfies $\hat{\cal G}_0(-\br_1,-\br_2;\omega_n)=\hat{\cal G}_0(\br_1,\br_2;\omega_n)$, and Eq. (\ref{tilde-K_ij}) yields $\tilde K_{ij}=0$. Therefore, the Goldstone modes corresponding to the phase fluctuations and elastic deformations are decoupled. This can also be understood using the following symmetry argument. 
Assuming a general three-dimensional displacement $\bm{u}=(u^1,u^2,u^3)$, one can introduce the displacement gradient tensor $\tilde\varepsilon^i_j=\partial u^i/\partial x^j$ (the usual strain tensor of the elasticity theory is the symmetric part of $\tilde\varepsilon^i_j$). For a uniform deformation of the LO phase, we have $u^1=u^2=0$, $u^3=\tilde\varepsilon^3_ix^i$, where $\beps=(\tilde\varepsilon^3_1,\tilde\varepsilon^3_2,\tilde\varepsilon^3_3)$ does not transform like a vector. In particular, $\beps$ remains invariant under inversion, while $\bvs$ changes sign, therefore the free energy cannot contain quadratic terms mixing $\beps$ and $\bvs$.

\section{LO phase in one dimension}
\label{sec: 1D}

As an application of the general formalism developed in the previous section, we consider the LO phase in a superconductor with a 1D band dispersion $\xi(k_z)=(k_z^2-k_F^2)/2m^*$, where $k_F$ is the Fermi wavevector and $m^*$ is the effective mass. In this case, the mean-field gap equations can be solved exactly, see Refs. \onlinecite{MN84} and \onlinecite{BP87}, using the formal similarity with the 1D Peierls problem considered in Ref. \onlinecite{BGK80}. 

\begin{figure}
    \includegraphics[width=9cm]{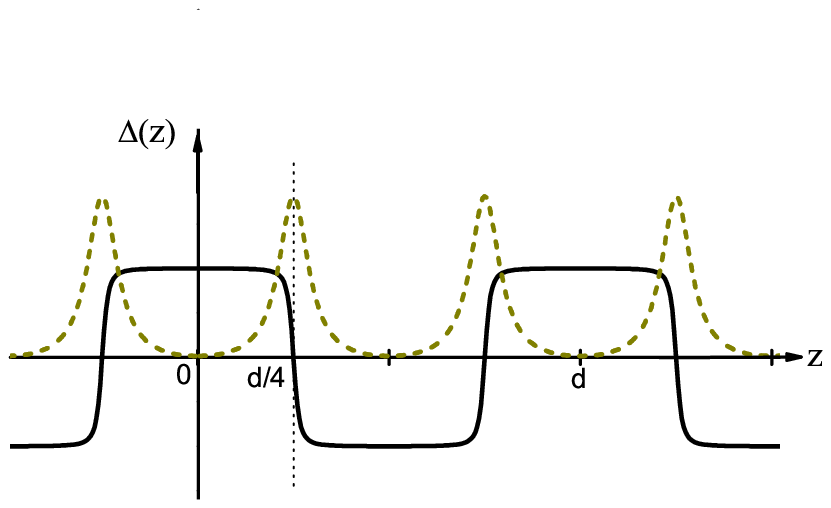}
    \caption{The order parameter (solid line) and the quasiparticle wavefunctions (dashed line) in the 1D LO phase.}
	\label{fig: LO-1D}
\end{figure}

Let us briefly summarize the relevant properties of the exact solution. Although the critical temperature decreases with the Zeeman field $h$, the superconductivity is not completely suppressed even at strong fields. The order parameter is described by the following expression:
\begin{equation}
\label{exact-Delta}
	\Delta(z)=\Delta_1\,\mathrm{cd}\left(\frac{\Delta_1}{v_Fk_1}z,k_1\right),
\end{equation}
where ``cd'' is a Jacobian elliptic function (Ref. \onlinecite{AS65}), $v_F=k_F/m^*$ is the Fermi velocity, and $\Delta_1$ and $k_1$ are parameters that depend on $T$ and $h$. The period of the LO structure is given by
\begin{equation}
\label{exact-period}
	d=\frac{4k_1\mathrm{K}(\gamma)}{1+k_1}\frac{v_F}{\Delta_1},
\end{equation}
where $\gamma=2\sqrt{k_1}/(1+k_1)$, and $\mathrm{K}(\gamma)$ is the complete elliptic integral of the first kind. At low temperatures, the LO structure resembles a soliton lattice, sketched in Fig. \ref{fig: LO-1D}. Note that our expression for the LO order parameter, Eq. (\ref{exact-Delta}), is shifted by a quarter-period compared to those in Refs. \onlinecite{MN84} and \onlinecite{BP87}, to make explicit the inversion symmetry $\Delta(-z)=\Delta(z)$. At zero temperature, one has $\Delta_1=\sqrt{k_1}\Delta_0$, where $\Delta_0$ is the BCS gap at $T=h=0$, while the parameter $\gamma$ as a function of $h$ is found from the equation $\mathrm{E}(\gamma)/\gamma=\pi h/2\Delta_0$, where $\mathrm{E}(\gamma)$ is the complete elliptic integral of the second kind. 

The energy density of the Goldstone modes, Eq. (\ref{E_LO}), takes the following form:
\begin{equation}
\label{E_LO-1D}
	\frac{\delta{\cal F}_{LO}}{\cal V}=\frac{1}{2}\rho_s v^2_{s,z}+\frac{1}{2}K\varepsilon^2_z,
\end{equation}
with the coefficients in the first and second terms characterizing the stiffness of the LO phase against the phase fluctuations and the deformations of the amplitude modulation, respectively. We have $\rho_s=\rho_1+\rho_2$, where
\begin{eqnarray}
\label{rho_s-1-2}
	\rho_1=\frac{m^2}{m^*}T\sum_n\frac{1}{L}\Tr(\hat\sigma_3\hat{\cal G}_0),\quad
	\rho_2=\frac{m^2}{m^{*,2}}T\sum_n\frac{1}{L}\Tr(\hat k_z\hat{\cal G}_0\hat k_z\hat{\cal G}_0),
\end{eqnarray}
and $K=K_1+K_2$, where
\begin{eqnarray}
\label{K-1-2}
	K_1=\frac{3}{m^*}T\sum_n\frac{1}{L}\Tr(\hat k_z^2\hat\sigma_3\hat{\cal G}_0),\quad
	K_2=\frac{1}{m^{*,2}}T\sum_n\frac{1}{L}\Tr(\hat k_z^2\hat\sigma_3\hat{\cal G}_0\hat k_z^2\hat\sigma_3\hat{\cal G}_0)
\end{eqnarray}
($L$ is the system length).

\begin{figure}
    \includegraphics[width=11cm]{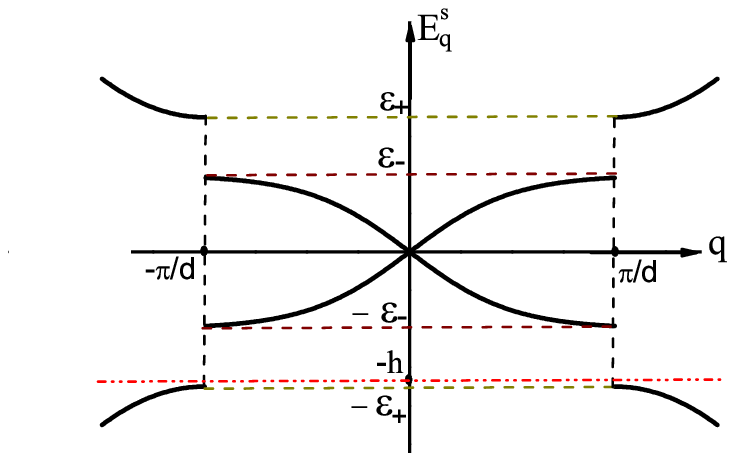}
    \caption{The quasiparticle excitation spectrum in the 1D LO phase, shown in the extended zone scheme (from Ref. \onlinecite{BGK80}).
	The Zeeman field $h$ is located inside the energy gap.}
	\label{fig: qp-spectrum}
\end{figure}

Since the order parameter varies slowly compared to the Fermi wavelength $k_F^{-1}$, one can use the quasiclassical, or Andreev, approximation. The wavefunctions can be represented in the form $e^{isk_Fz}\psi(z)$, where $s=\pm$ is the direction index, which labels the roots of the equation $\xi(k_z)=0$, and the slowly varying factors $\psi(z)$ are the eigenfunctions of the Andreev Hamiltonian:
\begin{equation}
\label{H-Andreev-s}
	\hat H_s=-isv_F\hat\sigma_3\frac{d}{dz}+\Delta(z)\hat\sigma_1.
\end{equation}
Due to the periodicity of $\Delta(z)$, the wavefunctions satisfy $\psi(z+d)=e^{iqd}\psi(z)$ and are characterized by the wavevector $q$. At given $s$ and $q$, the spectrum consists of two branches, labelled by the branch index $\nu=1,2$, with $E^s_{q,2}=-E^s_{q,1}$, see Fig. \ref{fig: qp-spectrum}. There are two energy gaps, located at $q=\pm\pi/d$, with the gap edges given by $\pm\epsilon_+$ and $\pm\epsilon_-$, where $\epsilon_-=\Delta_0\sqrt{1/\gamma^2-1}$ and $\epsilon_+=\Delta_0/\gamma$ (at $T=0$). It is easy to show that the Zeeman field is located inside the gap, i.e. $\epsilon_-<h<\epsilon_+$, and that $\epsilon_+-h<h-\epsilon_-$. 
The matrix Green's function can be written as follows:
\begin{equation}
\label{matrix-GF-1D-A}
	\hat{\cal G}_0(z,z';\omega_n)=\sum_s e^{isk_F(z-z')}
	\sum_{q,\nu}\frac{\psi^s_{q,\nu}(z)\psi^{s,\dagger}_{q,\nu}(z')}{i\omega_n-E^s_{q,\nu}-h}.
\end{equation}

The quasiclassical approximation cannot be directly applied to $\rho_1$, see Eq. (\ref{rho_s-1-2}), because of the contributions of the states far from the Fermi energy. To get around this difficulty, we subtract from and add to $\rho_1$ the corresponding expression in the normal state: $\rho_1=L^{-1}(m^2/m^*)T\sum_n\Tr(\hat\sigma_3\hat{\cal G}_{0,N})+\delta\rho_1$, where $\hat{\cal G}_{0,N}^{-1}(k_z,\omega_n)=i\omega_n-\xi(k_z)\hat\sigma_3-h$, while
$$
	\delta\rho_1=\frac{m^2}{m^*}T\sum_n\frac{1}{L}\left[\Tr(\hat\sigma_3\hat{\cal G}_0)-\Tr(\hat\sigma_3\hat{\cal G}_{0,N})\right]
$$
can be calculated quasiclassically. If the asymmetry of the quasiparticle density of states near the Fermi energy is neglected, then $\delta\rho_1=0$. The remaining expression is entirely determined by the normal-state properties and yields $\rho_1=(m^2/m^*)n$, where $n$ is the concentration of electrons.

The second contribution to the superfluid density, $\rho_2$, comes from the states in the vicinity of the Fermi energy and can therefore be calculated in the quasiclassical approximation.  
Substituting the Green's function (\ref{matrix-GF-1D-A}) in Eq. (\ref{rho_s-1-2}), differentiating only the fast exponentials, neglecting the integrals of functions that oscillate on the scale of $k_F^{-1}$, and using the orthonormality of the Andreev eigenstates, we obtain: 
\begin{equation}
\label{rho-2-gen}
	\rho_2=-m^2v_F^2\int dE\,N(E)\left[-\frac{\partial f(E+h)}{\partial E}\right],
\end{equation}
where $f(E)=(e^{\beta E}+1)^{-1}$ is the Fermi distribution function, and $N(E)=L^{-1}\sum_{sq\nu}\delta(E-E^s_{q,\nu})$
is the density of states of the Bogoliubov quasiparticles. At zero temperature, we have
$\rho_2=-m^2v_F^2N(-h)=0$, because $-h$ is located inside the gap, see Fig. \ref{fig: qp-spectrum}. At low temperatures, the corrections are exponentially small: $\rho_2(T)\propto e^{-E_g/T}$, where $E_g(h)=\epsilon_+-h=\Delta_0[1-2\mathrm{E}(\gamma)/\pi]/\gamma$. Putting all the pieces together, we find
\begin{equation}
\label{rho_s-1D-final}
	\rho_s(T)=\frac{m^2}{m^*}n-{\cal O}(e^{-E_g/T}).
\end{equation}
Thus the superfluid density at $T=0$ is the same as in a uniform BCS superconductor. At nonzero temperatures, it is reduced due to the thermally excited Bogoliubov quasiparticles, characterized by the energy gap $E_g$. 

Let us now turn to the calculation of the elastic modulus, see Eq. (\ref{K-1-2}). As in the case of $\rho_1$, the quasiclassical approximation cannot be applied directly to $K_1$, but one can write $K_1=(3/m^*L)T\sum_n\Tr(\hat k_z^2\hat\sigma_3\hat{\cal G}_{0,N})+\delta K_1$, where
$\delta K_1=(3k_F^2/m^2)\delta\rho_1=0$. Therefore,
\begin{equation}
	K_1=\frac{3}{m^*}T\sum_n\frac{1}{L}\sum_{k_z}k_z^2\,\tr[\hat\sigma_3\hat{\cal G}_{0,N}(k_z,\omega_n)]
	=\frac{2k_F^3}{\pi m^*},
\end{equation}
at zero temperature. Here we neglected the corrections of the order of $h/\epsilon_F$, where $\epsilon_F=k_F^2/2m^*$. The quasiclassical expression for $K_2$ reads
\begin{equation}
\label{K-2-gen}
	K_2=\frac{k_F^4}{m^{*,2}}\sum_s\frac{1}{L}
	\sum_{q_1\nu_1}\sum_{q_2\nu_2}|I^s_{q_1\nu_1,q_2\nu_2}|^2
	\frac{f(E^s_{q_1,\nu_1}+h)-f(E^s_{q_2,\nu_2}+h)}{E^s_{q_1,\nu_1}-E^s_{q_2,\nu_2}},
\end{equation}
where the matrix element is given by $I^s_{q_1\nu_1,q_2\nu_2}=\int_0^L dz\,\psi^{s,\dagger}_{q_1,\nu_1}(z)\hat\sigma_3\psi^s_{q_2,\nu_2}(z)$. 
The interband contribution is dominated by the ``vertical'' transitions near the gap edges at $q=\pm\pi/d$, where the matrix element vanishes for symmetry reasons. Keeping only the intraband terms in Eq. (\ref{K-2-gen}), we obtain:
\begin{equation}
 	K_2=-\frac{k_F^4}{m^{*,2}}\int dE\,\tilde N(E)\left[-\frac{\partial f(E+h)}{\partial E}\right],
\end{equation}
where $\tilde N(E)=L^{-1}\sum_{sq\nu}|I^s_{q\nu,q\nu}|^2\delta(E-E^s_{q,\nu})$. At zero temperature, we have $K_2\propto\tilde N(-h)=0$, while the low-temperature corrections are exponentially small and proportional to $e^{-E_g/T}$. Thus, the elastic modulus of the 1D LO state at low temperatures has the following form:
\begin{equation}
\label{K-1D-final}
	K(T)=\frac{2k_F^3}{\pi m^*}-{\cal O}(e^{-E_g/T}).
\end{equation}
Similarly to the superfluid density, it is reduced by the thermally activated Bogoliubov quasiparticles, until both $\rho_s$ and $K$ vanish at the critical temperature $T_c(h)$.

To conclude this section we note that the nonvanishing superfluid density $\rho_s$ implies that a uniform supercurrent can flow across the zeros of the order parameter amplitude. This can be understood using a simple phenomenological argument based on the Ginzburg-Landau theory for the LOFF state. Near the critical temperature, the lowest order terms in the gradient energy can be written as follows: ${\cal F}_{grad}=\int dz\,(\tilde K_2|D_z\Delta|^2+\tilde K_4|D_z^2\Delta|^2)$, where $D_z=-i\nabla_z$, and $\tilde K_2<0$, $\tilde K_4>0$, in order for the superconducting instability with a finite wavevector to occur.\cite{BK97} The supercurrent can be calculated in the usual way: $j_{s,z}=-c(\delta{\cal F}_{grad}/\delta A_z)$, after making the replacement $D_z\to -i\nabla_z+(2e/c)A_z$ in ${\cal F}_{grad}$, where $A_z$ is the vector potential and $e$ is the absolute value of the electron charge. We obtain:
\begin{equation}
	j_{s,z}=-4e\tilde K_2\im\left(\Delta^*\nabla_z\Delta\right)-4e\tilde K_4\im\left[(\nabla_z\Delta^*)\nabla_z^2\Delta-\Delta^*\nabla_z^3\Delta\right].
\end{equation}
Using the amplitude-phase representation of the order parameter: $\Delta(z)=|\Delta(z)|e^{i\theta(z)}$, and keeping only the lowest order terms in the superfluid velocity, we obtain:
$$
	j_{s,z}=-8em\left[\tilde K_2|\Delta|^2+2\tilde K_4((\nabla_z|\Delta|)^2-2|\Delta|\nabla_z^2|\Delta|)\right]v_{s,z}.
$$
Thus, the supercurrent is not just proportional to $|\Delta|^2$ and, therefore, does not vanish at the points where the order parameter amplitude has zeros.

\section{Conclusions}
\label{sec: conclusion}

We have derived general expressions for the energy of the Goldstone modes in the LOFF phases with one-dimensional periodicity, see Eq. (\ref{F_FF}) for the FF phase and Eq. (\ref{E_LO}) for the LO phase. While there is only one Goldstone mode in the FF state, corresponding to the fluctuations of the order parameter phase, there are additional, ``elastic'', modes in the LO state, corresponding to the compression and tilting of the order parameter amplitude modulation. Our approach, which is based on the transformation of slow deformations of the order parameter into small corrections to the Bogoliubov-de Gennes Hamiltonian, allows one to treat the fluctuation effects perturbatively. It can be easily generalized to more complicated LOFF structures. 

Using the exact solution for the excitation spectrum in a one-dimensional LO phase, we calculated the stiffness coefficients for the phase and elastic Goldstone modes. At $T=0$ the phase stiffness (the superfluid density) has the same form as in a uniform superconductor, while at low temperatures the corrections are exponential, with a field-dependent energy gap.\\  

This work was supported by a Discovery Grant from the Natural Sciences and Engineering Research Council of Canada.

\appendix

\section{Counting Goldstone modes in the LOFF state}
\label{sec: Goldstones}

In the vicinity of the critical temperature $T_c(h)$, the mean-field free energy of a LOFF superconductor can be expanded in powers of the order parameter as follows:
\begin{eqnarray}
\label{F}
    {\cal F}=\sum\limits_{\bq} A(\bq)\Delta^*_{\bq}\Delta_{\bq}+
    \sum\limits_{\bq_{1,2,3,4}}B(\bq_1,\bq_2,\bq_3,\bq_4)\Delta^*_{\bq_1}\Delta^*_{\bq_2}\Delta_{\bq_3}\Delta_{\bq_4}
    \delta_{\bq_1+\bq_2,\bq_3+\bq_4}.
\end{eqnarray}
Since the full momentum dependence of $A$ and $B$ is retained, Eq. (\ref{F}) can be used close to the second-order phase transition line at all $h$. The coefficient $A(\bq)$ changes sign at the critical temperature of the superconducting instability with the wave vector $\bq$. The explicit form of the functions $A$ and $B$ depends on the microscopic details.

The order parameter in the LOFF state close to the critical temperature can be represented in the following form:
\begin{equation}
\label{LOFF OP}
    \Delta(\br)=\sum_{i=1}^{N_Q}\Delta_i e^{i\bQ_i\br},
\end{equation}
where $\Delta_i=|\Delta_i|e^{i\phi_i}$ are complex coefficients, $\bQ_i$ are the positions of the degenerate minima of $A(\bq)$, and $N_Q$ is the number of the minima. In order to find the order parameter components $\Delta_i$, which determine the spatial structure of the LOFF phase,
one needs to evaluate the free energy (\ref{F}) for the order parameter (\ref{LOFF OP}). We obtain: ${\cal F}/{\cal V}=F_2+F_4$, where 
$F_2=A_0\sum_i|\Delta_i|^2$, with $A_0=A(\bQ_i)$ (same for all $i$), and
\begin{equation}
\label{F4-gen}
    F_4=\sum\limits_{ijkl}B_{ijkl}\Delta_i^*\Delta_j^*\Delta_k\Delta_l,
\end{equation}
with $B_{ijkl}=B(\bQ_i,\bQ_j,\bQ_k,\bQ_l)\delta_{\bQ_i+\bQ_j,\bQ_k+\bQ_l}$. In addition to the momentum conservation and the conditions $B_{ijkl}=B_{jikl}=B_{ijlk}$, the
coefficients $B_{ijkl}$ must also satisfy certain symmetry-imposed constraints, which can be obtained from the requirement that 
the order parameter transforms like a scalar function under an arbitrary operation $g$ from the point group $\mathbb{G}$ of the crystal, i.e. $\Delta(\br)\to\Delta(g^{-1}\br)$. Inserting here the expansion (\ref{LOFF OP}) and taking into account that the set of $\bQ_i$'s is invariant under all operations from $\mathbb{G}$, we see that the action of $g$ on the set of $\Delta_i$'s is equivalent to a permutation $P(g)$. Therefore, Eq. (\ref{F4-gen}) must remain invariant under $P(g)$ for all $g$.

Let us illustrate the above statements using as an example a tetragonal crystal with $\mathbb{G}=\mathbf{D}_{4h}$. The point group has 16 elements, therefore
there can be as many as $N_Q=16$ degenerate minima of $A(\bq)$. The general case is clearly untreatable, so we consider just two simplest cases: $N_Q=2$, in which
the deepest minima of $A$ are located at
$\bQ_{1,2}=\pm\bQ$, where $\bQ=Q\hat z$; and $N_Q=4$, in which the minima occupy four high-symmetry points in the basal plane:
$\bQ_{1,2}=\pm Q\hat x$, $\bQ_{3,4}=\pm Q\hat y$.

\underline{$N_Q=2$}. The order parameter has the form $\Delta(\br)=\Delta_1e^{iQz}+\Delta_2e^{-iQz}$. The free energy must be invariant under the action of the generators of $\mathbf{D}_{4h}$, i.e. rotations $C_{4z}$ and $C_{2x}$, and also inversion $I$. Due to the momentum conservation, the nonzero coefficients in Eq. (\ref{F4-gen}) are
$B_{1111}=B(\bQ,\bQ,\bQ,\bQ)$, $B_{1212}=B_{2112}=B_{1221}=B_{2121}=B(\bQ,-\bQ,\bQ,-\bQ)$, and $B_{2222}=B(-\bQ,-\bQ,-\bQ,-\bQ)$. Since the inversion operation interchanges $\Delta_1$ and $\Delta_2$, we have $B_{1111}=B_{2222}$, and Eq. (\ref{F4-gen}) takes the following form:
\begin{equation}
\label{F4 N2}
    F_4=\beta_1(|\Delta_1|^4+|\Delta_2|^4)+\beta_2|\Delta_1|^2|\Delta_2|^2,
\end{equation}
where $\beta_1=B_{1111}$ and $\beta_2=4B_{1212}$. The free energy does not depend on the phases of $\Delta_{1,2}$, therefore there might be up to two Goldstone modes in the LOFF state, corresponding to the fluctuations of either the overall phase of the superconducting order parameter or the relative phase of $\Delta_{1,2}$. Expression (\ref{F4 N2}) is positive definite if $\beta_1>0$, $\beta_2>-2\beta_1$. If $\beta_2<2\beta_1$, then the minimum energy is achieved for $|\Delta_1|=|\Delta_2|=\Delta_0$, which corresponds to the LO phase with $\Delta(\br)=\Delta_0e^{i\theta}\cos(Qz+\varphi)$. The relative phase of $\Delta_{1,2}$ describes a uniform translation of the order parameter. On the other hand, if $\beta_2>2\beta_1$, then $|\Delta_2|=0$ or $|\Delta_1|=0$, corresponding to one of the two degenerate FF phases,
$\Delta(\br)=\Delta_0e^{i\theta}e^{iQz}$, or $\Delta(\br)=\Delta_0e^{i\theta}e^{-iQz}$.

\underline{$N_Q=4$}. The order parameter has the form $\Delta(\br)=\Delta_1e^{iQx}+\Delta_2e^{-iQx}+\Delta_3e^{iQy}+\Delta_4e^{-iQy}$.
Under $C_{4z}$, we have $\Delta_{1,2,3,4}\to\Delta_{3,4,2,1}$, while under $C_{2x}$ and $I$, $\Delta_1\leftrightarrow\Delta_2$ and $\Delta_3\leftrightarrow\Delta_4$. The nonzero coefficients in Eq. (\ref{F4-gen}) are as follows: $B_{1111}=B_{2222}=B_{3333}=B_{4444}=\beta_1$, $B_{1212}=B_{3434}=\beta_2/4$, $B_{1313}=B_{1414}=B_{2323}=B_{2424}=\beta_3/4$, and $B_{1234}=B_{3412}=\beta_4/4$, therefore,
\begin{eqnarray}
\label{F4 N4}
	F_4&=&\beta_1(|\Delta_1|^4+|\Delta_2|^4+|\Delta_3|^4+|\Delta_4|^4)+\beta_2(|\Delta_1|^2|\Delta_2|^2+|\Delta_3|^2|\Delta_4|^2)\nonumber\\
	&&+\beta_3(|\Delta_1|^2+|\Delta_2|^2)(|\Delta_3|^2+|\Delta_4|^2)+\beta_4(\Delta_1^*\Delta_2^*\Delta_3\Delta_4+\Delta_1\Delta_2\Delta_3^*\Delta_4^*).
\end{eqnarray}
The last term leads to a phase locking of the order parameter components: at the energy minimum the phases of $\Delta_i$ are subject to the constraint $\phi_1+\phi_2-\phi_3-\phi_4=0$ or $\pi$, depending on the sign of $\beta_4$. Therefore, there can be up to three Goldstone modes, one of which corresponds to the fluctuations of the overall phase of $\Delta(\br)$.

\end{document}